\documentclass[twocolumn,showpacs,preprintnumbers,amsmath,amssymb,superscriptaddress]{revtex4-1}

\usepackage{graphicx}
\usepackage{dcolumn}
\usepackage{bm}
\usepackage{tabularx}
\usepackage{makecell}
\usepackage{braket}
\usepackage{lipsum}
\usepackage{ulem}
\usepackage{amsmath,amssymb} 
\newcolumntype{M}{>{\centering\arraybackslash}m{1.85cm}}
\usepackage[export]{adjustbox}
\usepackage{longtable}
\usepackage{float}
\usepackage{xcolor}
\usepackage{color}   
\usepackage{hyperref}
\hypersetup{
	colorlinks=true, 
	linkcolor=blue,  
}

\makeatletter
\newcommand{\colorcaption}[2][]{%
	\begingroup%
	\renewcommand{\@caption@fignum@sep}{ (Color online). }%
	\caption[#1]{#2}%
	\endgroup%
}
\makeatother

\begin{document}
	\title{Systematic shell-model analysis of $2\nu\beta\beta$ decay of $^{76}$Ge and $^{96}$Zr to the ground and excited states of $^{76}$Se and $^{96}$Mo}
	
	\author{ Deepak Patel}
	\address{Department of Physics, Indian Institute of Technology Roorkee, Roorkee 247667, India}
	\author{Praveen C. Srivastava}
	\email{Corresponding author: praveen.srivastava@ph.iitr.ac.in}
	\address{Department of Physics, Indian Institute of Technology Roorkee, Roorkee 247667, India}
	\author{Jouni Suhonen}
	\address{University of Jyvaskyla, Department of Physics, P.O. Box 35 (YFL), FI-40014, 
University of Jyvaskyla, Finland}
        \address{International Centre for Advanced Training and Research in Physics (CIFRA),P.O. Box MG12, 077125 Bucharest-M\u{a}gurele, Romania}
	
	\date{\hfill \today}
	
	
	\begin{abstract}
		
		In this work, we have studied the $2\nu\beta\beta$ decay of $^{76}$Ge and $^{96}$Zr isotopes utilizing large-scale shell-model calculations. The GWBXG effective interaction has been employed in the calculation of $2\nu\beta\beta$-decay nuclear matrix elements (NMEs). We have tested the effective interaction by comparing the predicted spectroscopic properties, such as energy spectra and transition probabilities, with the available experimental data. The variation of cumulative NMEs with respect to the $1^+$ state energies of the intermediate nucleus is also studied, corresponding to $0^+_{\rm g.s.}\rightarrow0^+_{\rm g.s.}$, $0^+_{\rm g.s.}\rightarrow0^+_{2}$, and $0^+_{\rm g.s.}\rightarrow2^+_{1}$ transitions between the parent and grand-daughter nuclei. The effective values of axial-vector coupling strength ($g_A^{\rm eff}$) are calculated using the predicted NMEs and experimental half-lives for $0^+_{\rm g.s.}\rightarrow0^+_{\rm g.s.}$ transitions. The extracted half-lives for $0^+_{\rm g.s.}\rightarrow0^+_{2}$, and $0^+_{\rm g.s.}\rightarrow2^+_{1}$ transitions using the shell-model predicted NMEs are consistent with the recent experimental data. The comparison of the shell-model predicted NMEs with previous NMEs available in the literature is discussed. Also, the computed branching ratios for the $2\nu\beta\beta$ decay of $^{76}$Ge and both the $2\nu\beta\beta$ and single-$\beta$ decay of $^{96}$Zr are reported corresponding to the calculated $g_A^{\rm eff}$ values.
		
	\end{abstract}
	
	
	\maketitle
	
	\section{Introduction}
	
	Double $\beta$ decay (DBD) is a rare second-order weak-interaction process wherein a nucleus undergoes a simultaneous conversion of two neutrons into two protons or vice versa \cite{Deppisch, GERDA, Agostini, Ackerman, Adams, Arnold3, Elliott}. This phenomenon, first proposed by Goeppert-Mayer in 1935 as a nuclear disintegration \cite{Mayer}, plays a significant role in nuclear and particle physics. DBD occurs primarily in isotopes where single $\beta$ decay is energetically forbidden or highly suppressed. This process can occur through two decay modes: two-neutrino double $\beta$ decay ($2\nu\beta\beta$) and neutrino-less double $\beta$ decay ($0\nu\beta\beta$). In the realm of double $\beta$ decay studies, $2\nu\beta\beta$ decay offers a valuable assessment of the standard model. It also confirms the validity of the weak nuclear forces and the existence of neutrinos as weakly interacting particles. The double-$\beta$-minus mode of this decay process, relevant for the present studies, can be written as: $(A,Z)\rightarrow(A,Z+2)+2e^-+2\bar{\nu}_e$. In previous years, the $2\nu\beta\beta$-decay process has been observed in different mass regions of the nuclear chart for several nuclei. The study of the $2\nu\beta\beta$ decay mode also provides natural background components in $0\nu\beta\beta$ measurements \cite{Chandra,Ejiri}. In the case of the $0\nu\beta\beta$ decay process, which is not observed yet, neutrino emission does not occur, thus violating 
lepton-number conservation \cite{Rodriguez, Shimizu1}. Determination of the half-lives of $2\nu\beta\beta$ and $0\nu\beta\beta$ decaying nuclei imparts crucial details about the decay rates and lifetimes of nuclei involved in the nucleosynthesis phenomenon \cite{Zuber}. In the present study, we consider only the $2\nu\beta\beta$ decay process.
	
Several experiments have been performed for the accurate half-life estimation of the $ 2\nu\beta\beta$-decaying nuclei by different groups in the last few decades. In the NEMO-3 experiment, Arnold \textit{et al.} \cite{Arnold} measured the half-life for $2\nu\beta\beta$ decay of $^{48}$Ca to the ground state (g.s.) of $^{48}$Ti as $T_{1/2}^{2\nu}=[6.4^{+0.7}_{-0.6}\text{(stat)}^{+1.2}_{-0.9}\text{(syst)}]\times10^{19}$ yr. Recently, Agostini \textit{et al.} \cite{Agostini} extracted the half-life of $^{76}$Ge for $2\nu\beta\beta$ decay to the g.s. of $^{76}$Se as $(2.022\pm0.018_{\text{stat}}\pm0.038_{\text{syst}})\times 10^{21}$ yr from the GERmanium Detector Array (GERDA). From another NEMO-3 experiment, Arnold \textit{et al.} \cite{Arnold1} studied the $2\nu\beta\beta$ decay in $^{82}$Se to the g.s. of $^{82}$Kr; they have measured the half-life as $[9.39\pm0.17\text{(stat)}\pm0.58\text{(syst)}]\times10^{19}$ yr and extracted the corresponding nuclear matrix element $|M_{2\nu}|=0.0498\pm0.0016$ using the value $g_{\rm A}=1.27$ for the weak axial coupling. In the $A=90$ mass region, $^{96}$Zr is a well-known candidate of $2\nu\beta\beta$ decay. Using the NEMO-3 detector, the obtained half-life of $^{96}$Zr in $2\nu\beta\beta$ decay to the g.s. of $^{96}$Mo is $T_{1/2}^{2\nu}=[2.35\pm0.14\text{(stat)}\pm0.16\text{(syst)}]\times10^{19}$ yr and the extracted NME using the above half-life and $g_{\rm A}=1.25$ is $M_{2\nu}=0.049\pm0.002$ \cite{Argyriades}. Another candidate of $2\nu\beta\beta$ decay, $^{100}$Mo has one of the largest decay energies, $Q_{\beta\beta}=3034.36(17)$ keV \cite{NNDC_qcal} and the shortest half-life \cite{Barabash}. Recently, Armengaud \textit{et al.} \cite{Armengaud} determined the g.s.-to-g.s. decay half-life of $^{100}$Mo very precisely as $T_{1/2}^{2\nu}=[7.12^{+0.18}_{-0.14}\text{(stat)}\pm 0.10\text{(syst)}]\times10^{18}$ yr using the CUPID-Mo detection technology. In the Aurora experiment, the half-life of the $2\nu\beta\beta$ transition of $^{116}$Cd to the $0^+_{\rm g.s.}$ of $^{116}$Sn was measured with the highest accuracy as $T_{1/2}^{2\nu}=(2.63^{+0.11}_{-0.12})\times 10^{19}$ yr \cite{Barabash2}. In another work, the half-lives of two other $2\nu\beta\beta$ decaying candidates, $^{128,130}$Te, were proposed as $(2.41\pm0.39)\times 10^{24}$ and $(9.0\pm1.4)\times 10^{20}$ yr, respectively, using geological Te specimens \cite{Meshik}. In the NEXT Collaboration, Novella \textit{et al.} \cite{Novella} extracted the $2\nu\beta\beta$ half-life of $^{136}$Xe as $[2.34^{+0.80}_{-0.46}(\text{stat})^{+0.30}_{-0.17}(\text{syst})]\times 10^{21}$ yr for the g.s. transition. After the nucleus $^{48}$Ca, $^{150}$Nd has the second largest $Q_{\beta\beta}$ value (3371.38 keV \cite{NNDC_qcal}), making it an intriguing nucleus for studying both $2\nu\beta\beta$ and $0\nu\beta\beta$ decay. Recently, Arnold \textit{et al.} \cite{Arnold2} reported findings on the search for $0\nu\beta\beta$ decay and measured the $2\nu\beta\beta$ decay half-life of $^{150}$Nd as $T_{1/2}^{2\nu}=[9.34\pm0.22\text{(stat)}^{+0.62}_{-0.60}\text{(syst)}]\times 10^{18}$ yr for the g.s.-to-g.s transition in the NEMO-3 Collaboration.

	The cumulative nuclear matrix elements $(M_{2\nu})$ for $2\nu\beta\beta$ decay should show a significant contribution from the Gamow-Teller giant resonance (GTGR) region. Nuclear models like proton-neutron quasi particle random-phase approximation ($pn$QRPA) take the GTGR region into account in a realistic way \cite{Suhonen1, Terasaki}. However, for the nuclear shell model the high energy in the GTGR region remains a challenge and could only partly be described by two earlier shell-model calculations in the case of $^{48}$Ca \cite{Kostensalo1, Horoi}. In the past, due to huge dimensions, it was difficult to perform large-scale shell-model calculations for $2\nu\beta\beta$ decay studies in the medium to heavier mass region. At present, it is possible to perform shell-model calculations in extended valence spaces, including the relevant shell-model configurations, due to significant progress in configuration mixing using novel approaches and new advancements in the computational facilities. All this could help us describe the GTGR region realistically within the shell-model framework in the near future.

    We have studied the $2\nu\beta\beta$ decay in several nuclei in our previous work \cite{Patel1}, excluding $^{76}$Ge and $^{96}$Zr. Now, motivated by the recent experimental data \cite{Agostini, Finch, Pritychenko} on the $2\nu\beta\beta$-decaying nuclei $^{76}$Ge and $^{96}$Zr, we have performed large-scale shell-model calculations for studying the behavior of cumulative NMEs with respect to the $1^+$ state energies in the intermediate nucleus and extracting the $2\nu\beta\beta$ half-lives of these two nuclei [using Eqs. (\ref{eq1}) and (\ref{eq2}) below]. The contributions of the $1^+$ states of the intermediate nucleus have been included up to the saturation level of NMEs. The present work represents the most comprehensive set of shell-model results thus far for the various $\beta$ and/or $2\nu\beta\beta$ transitions in $^{76}$Ge and $^{96}$Zr.

    This paper is organized as follows. Section \ref{section2} provides a concise introduction to the theoretical framework used in our calculations. Section \ref{section3} presents the outcomes of the shell-model analyses and discussion regarding the energy spectra, transition probabilities, NMEs, effective axial-vector coupling strength, extracted half-lives, level density, and branching ratios. Finally, we summarize our findings and draw conclusions in Sec. \ref{section4}.

	\section{Theoretical Framework} \label{section2}
	
	\subsection{Half-life}
	
\begin{figure}
		\includegraphics[width=88mm]{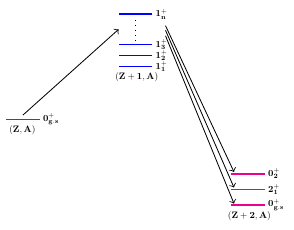}
		
		\caption{\label{DBD} Nuclear level and decay scheme for $2\nu\beta\beta$-minus decay.}
	\end{figure}

	We can express the half-life for the $2\nu\beta\beta$ decay as follows
	
	\begin{equation} \label{eq1}
		(T_{1/2}^{2\nu})^{-1}=G^{2\nu}(g_{\rm A}^{\rm eff})^4|M_{2\nu}|^2.
	\end{equation}
	Here, $G^{2\nu}$ denotes the phase-space factor \cite{Neacsu, Stoica, Kotila, Kotila1} and $g_{\rm A}^{\rm eff}$ is the effective axial-vector coupling strength \cite{Suhonen}. $M_{2\nu}$ is the nuclear matrix element (NME) for $2\nu\beta\beta$ decay and first derived by Tomoda \textit{et al.} \cite{Tomoda}. We can express it by the following expression \cite{Kostensalo}:
	
	\begin{equation} \label{eq2}
		M_{2\nu}=\sum_{N}\frac{\langle J^+||\sigma\tau^{\pm}||1_N^+\rangle \langle 1_N^+||\sigma\tau^{\pm}||0_{\rm g.s.}^{(i)}\rangle}{\sqrt{J+1}([\frac{1}{2}Q_{\beta\beta}+E(1^+_N)-M_i]/m_e+1)^k},
	\end{equation}
	where $m_e$ represents the rest mass of the electron; $E(1^+_N)-M_i$ corresponds the energy difference between the $N$th intermediate $1^+$ state and the ground state (g.s.) of the initial nucleus; $0_{\rm g.s.}^{(i)}$ stands for the ground state of the initial nucleus; $\sigma$ is the Pauli spin operator; $\tau^-(\tau^+)$ is the isospin lowering (raising) operator. $Q_{\beta\beta}$ ($Q$ value) is the energy released in the decay. $\langle J^+||\sigma\tau^{\pm}||1_N^+\rangle$ and $\langle 1_N^+||\sigma\tau^{\pm}||0_{\rm g.s.}^{(i)}\rangle$ denote the reduced Gamow-Teller (GT) matrix elements and $J$ represents the spin state of the final nucleus. Here, $k=1$ for $J=0$ and $k=3$ for $J=2$ \cite{Doi}. The nuclear energy and decay scheme for $2\nu\beta\beta$-minus decay is shown in Fig. \ref{DBD}. Here, it is appropriate to note that in the shell-model philosophy, the $g_{\rm A}^{\rm eff}$ of Eq. (\ref{eq1}) stems from the effective renormalization of the spin-isospin operator $\sigma\tau^\pm$ in the NME of Eq. (\ref{eq2}). After moving $g_{\rm A}^{\rm eff}$ in front of the squared NME of Eq. (\ref{eq1}), the $M_{2\nu}$ of Eq. (\ref{eq2}) is independent of the effective axial coupling. This contrasts the philosophy of $pn$QRPA where the computed NME depends on $g_{\rm A}^{\rm eff}$ through the renormalization of the model Hamiltonian by using the particle-particle interaction parameter $g_{\rm pp}$, Ref. \cite{Suhonen2017} being a tangible example of this philosophy.

	\subsection{Shell-model Hamiltonian}

	We can express the nuclear shell-model Hamiltonian in terms of single-particle energies and two-nucleon interactions as \cite{Patel2}
	
	\begin{multline}
		H = T+V = \sum_{\alpha}\epsilon_{\alpha}\hat{N}_\alpha+\sum_{\alpha \le \beta, \gamma \le \delta, J, M}\langle j_\alpha j_\beta |V|j_\gamma j_{\delta} \rangle_J \times \\
		O_{\alpha,\beta}^{\dagger}(JM)O_{\gamma,\delta}(JM),
	\end{multline}
	where $\alpha=\{nljt\}$ stand for the single-particle orbitals with $n$ being the number of nodes of the wave function, $l$ the orbital angular momentum, $j$ the total angular momentum, and $t=1/2$ the isospin. The $\epsilon_{\alpha}$ represent the corresponding single-particle energies. $\hat{N}_{\alpha}=\sum_{m_{\alpha}}c^{\dagger}_{\alpha,m_{\alpha}}c_{\alpha,m_{\alpha}}$ denotes the particle-number operator. The symbol $\langle j_\alpha j_\beta |V|j_\gamma j_\delta \rangle_{J}$ is the two-body interaction matrix element. $O_{\alpha,\beta}^{\dagger}(JM)(O_{\gamma,\delta}(JM))$ corresponds to the fermion pair creation (annihilation) operator and is written as
	
	\begin{multline}
	O_{\alpha,\beta}^{\dagger}(JM)=\frac{1}{\sqrt{1+\delta_{\alpha \beta}}}\sum_{m_{\alpha},m_{\beta}}\langle j_{\alpha} m_{\alpha} j_{\beta} m_{\beta} |JM\rangle \times \\ c^{\dagger}_{\alpha,m_{\alpha}} c^{\dagger}_{\beta,m_{\beta}}.
	\end{multline}

	We have employed the GWBXG effective shell-model Hamiltonian in our calculations, where its mean-field part consists of the $0f_{5/2}$, $1p_{3/2}$, $1p_{1/2}$, and $0g_{9/2}$ proton orbitals, as well as $1p_{1/2}$, $0g_{9/2}$, $0g_{7/2}$, $1d_{5/2}$, $1d_{3/2}$, and $2s_{1/2}$ neutron orbitals. This interaction is a composition of different interactions. The initial 974 two-body matrix elements (TBMEs) are derived from the bare $G$-matrix of the H7B potential \cite{Hosaka}. The bare $G$ matrix is not reasonable due to the space truncation, and the interaction should be renormalized by considering the core-polarization effects. Here, the present $G$-matrix effective interaction is tuned by further modification in matrix elements using fitted interactions: The 65 TBMEs for proton orbitals are replaced with the effective values reported in Ref. \cite{Ji}. The TBMEs connecting the $\pi(1p_{1/2}, 0g_{9/2})$ and the $\nu(1d_{5/2}, 2s_{1/2})$ orbitals are replaced by those from the Ref. \cite{Gloeckner}. Further, Serduke \textit{et al.} \cite{Serduke} replaced the TBMEs between the $\pi(1p_{1/2}, 0g_{9/2})$ and the $\nu(1p_{1/2}, 0g_{9/2})$ orbitals.
	
	It is difficult to calculate all possible $1^+$ states of intermediate nuclei in the full model space due to the involved huge dimensions. Thus, we have employed truncation on neutron orbitals above $N=50$ for calculating the $2\nu\beta\beta$-decay NME in $^{76}$Ge. Here, we have performed $1p-1h$ excitation across $N=50$ to the $0g_{7/2}$, $1d_{5/2}$, $1d_{3/2}$, and $2s_{1/2}$ neutron orbitals. For the calculation of the $2\nu\beta\beta$-decay NME in $^{96}$Zr, we have applied the truncation, where the considered proton and neutron partitions belong to the $\pi[(0f_{5/2}1p_{3/2})^{0-10}(1p_{1/2}0g_{9/2})^{0-4}]$ and $\nu[(1p_{1/2}0g_{9/2})^{12-12}(0g_{7/2}1d_{5/2}1d_{3/2}2s_{1/2})^{0-20}]$ configurations, respectively. The shell-model codes NUSHELLX \cite{Nushellx} and KSHELL \cite{Kshell} have been used in the diagonalization of the shell-model Hamiltonian matrices. Mainly, the NUSHELLX code is utilized to calculate the NMEs for the Gamow-Teller transitions, and the KSHELL code is used for the calculation of the energy spectra and transition probabilities of parent and grand-daughter nuclei.

	\section{Results and Discussion} \label{section3}
	
	In this section, we present the results obtained from shell-model calculations. First, we discuss the theoretical and experimental energy spectra and the corresponding quadrupole-reduced transition probabilities [$B(E2)$] for the parent and granddaughter nuclei of interest in the $2\nu\beta\beta$-decay study. We also show the variation of the shell-model calculated cumulative NME as a function of the excitation energy of the $1^+$ states of the intermediate nucleus. We estimate the $g_{\rm A}^{\rm eff}$ value utilizing the calculated NMEs for $0^+_{\rm g.s.}\rightarrow0^+_{\rm g.s.}$ $2\nu\beta\beta$ transitions and measured half-lives. Using the final predicted NMEs for $0^+_{\rm g.s.}\rightarrow0^+_2$, and $0^+_{\rm g.s.}\rightarrow 2^+_1$ transitions, we extract the $2\nu\beta\beta$-decay half-lives of $^{76}$Ge and $^{96}$Zr. We also compare our calculated NMEs with the previously available NMEs in the literature and the shell-model calculated level density of $1^+$ states in $^{76}$As and $^{96}$Nb with the back-shifted Fermi gas model (BFM). Lastly, we estimate the branching ratios for the $2\nu\beta\beta$ decay of $^{76}$Ge and for both the single-$\beta$ and $2\nu\beta\beta$ decay modes of $^{96}$Zr.

	
	\begin{figure}
		\includegraphics[width=42.5mm]{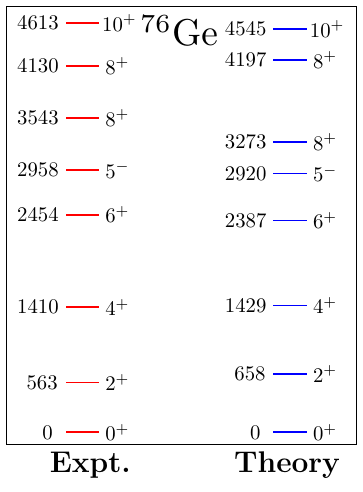}
		\includegraphics[width=42.5mm]{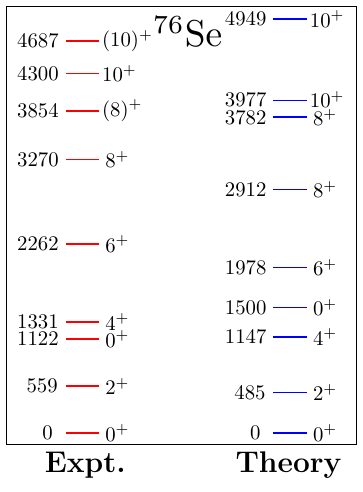}
		\caption{\label{76Ge_spectra} Comparison between experimental \cite{ENSDF} and shell-model calculated low-lying states in $^{76}$Ge and $^{76}$Se.}
	\end{figure}

	\subsection{Energy spectra and electromagnetic properties}

\begin{table}
		\centering
		\caption{Comparison of theoretical and experimental $B(E2)$ strengths [in Weisskopf units (W.u.)] \cite{NNDC_NUDAT} in the parent and grand-daughter nuclei of $2\nu\beta\beta$ decay. The effective charges are taken as $e_\pi=1.6e,e_\nu=0.8e$ \cite{Honma}.}
		\begin{ruledtabular}
			\begin{tabular}{cccccc}
				
				
				Isotope	&	$J_i^{\pi} \rightarrow J_f^{\pi}$   & Theory  &	Expt. 	 \\
				
				\hline \\
				
				$^{76}$Ge	&	$2^{+}_{1}$$\rightarrow$$0^{+}_{1}$   & 17.3 & 28.81(21) \\
				&	$4^{+}_{1}$$\rightarrow$$2^{+}_{1}$   & 22.8 & 36.5(8) \\		
				
				\hline \\
				
				$^{76}$Se	&	$2^{+}_{1}$$\rightarrow$$0^{+}_{1}$   & 28.9 & 45.1(+12-6) \\
				&	$4^{+}_{1}$$\rightarrow$$2^{+}_{1}$   & 42.4 & 71.1(14) \\
				&	$6^{+}_{1}$$\rightarrow$$4^{+}_{1}$   & 48.2 & 72.7(+68-58) \\
				&	$8^{+}_{1}$$\rightarrow$$6^{+}_{1}$   & 50.3 & 82(+21-14) \\
				&	$10^{+}_{1}$$\rightarrow$$8^{+}_{1}$  & 53.9 & 52(9) \\		
				\hline \\
				
				$^{96}$Zr	&	$2^{+}_{1}$$\rightarrow$$0^{+}_{1}$   & 2.4 & 2.3(3) \\
				&   $3^{+}_{1}$$\rightarrow$$2^{+}_{1}$   &  0.005  & 0.1(+3-1) \\
				&   $8^{+}_{1}$$\rightarrow$$6^{+}_{4}$   &  1.5  & 1.38(11) \\
				
				\hline \\
				
				$^{96}$Mo	&	$2^{+}_{1}$$\rightarrow$$0^{+}_{1}$   & 10.2 & 20.7(4) \\
				&   $4^{+}_{1}$$\rightarrow$$2^{+}_{1}$   & 11.8 & 41(7) \\
				&   $3^{+}_{1}$$\rightarrow$$2^{+}_{1}$   & 0.037 & $<1.3$ \\
				&   $6^{+}_{1}$$\rightarrow$$4^{+}_{1}$   & 9.3 & $<2.9\times10^{2}$ \\
				
			\end{tabular}
		\end{ruledtabular}
		\label{BE2}
	\end{table}

	In Fig. \ref{76Ge_spectra}, we depict the comparison between the shell-model and experimental low-energy states of the $^{76}$Ge and $^{76}$Se isotopes. It is clear from the spectra that our calculated states are in quite good agreement with the experimental data for both isotopes. In $^{76}$Ge, the shell-model predicted positive-parity yrast states (g.s. band) up to $8^+_1$ are characterized by mainly [$\pi(f_{5/2}^4)\otimes\nu(g_{9/2}^6)$] configuration, whereas the $10^+_1$ state stems from $[\pi(f_{5/2}^2p_{3/2}^2)\otimes\nu(g_{9/2}^6)\approx25.3\%]$ and [$\pi(f_{5/2}^4)\otimes\nu(g_{9/2}^6)\approx21.8\%$] configurations. Here, the large energy gap between the $8^+_1$ and $10^+_1$ states compared to the other two consecutive yrast states may be caused by the excitation of two protons from the $\pi(f_{5/2})$ orbital to the $\pi(p_{3/2})$ orbital in the dominant configuration of the $10^+_1$ state. Contrariwise, the yrast $0^+_{\rm g.s.}-10^+_1$ states (g.s. band) in $^{76}$Se are described by same dominant configuration $\pi(f_{5/2}^2p_{3/2}^2g_{9/2}^2)\otimes\nu(g_{9/2}^4)$. The excited states in the yrast band of $^{76}$Ge and $^{76}$Se show a pronounced collective behavior. The experimental data for $B(E2)$ transitions in the yrast band of $^{76}$Ge are available only for $2^+_1-0^+_1$ and $4^+_1-2^+_1$ transitions. As reported in Table \ref{BE2}, except for the $10^+_1-8^+_1$ transition in $^{76}$Se, our calculated $B(E2)$ transitions for both isotopes are slightly smaller but consistent with the experimental data.

       \begin{figure}
		\includegraphics[width=42.5mm]{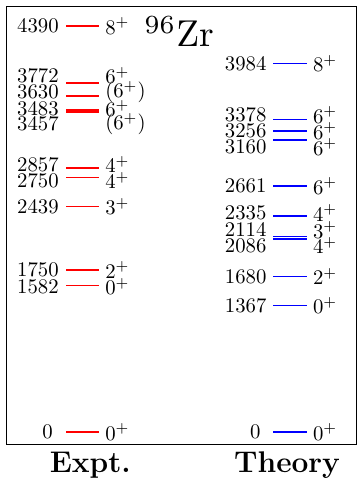}
		\includegraphics[width=42.5mm]{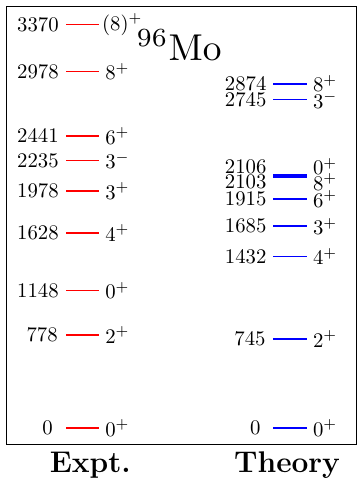}
		\caption{\label{96Zr_spectra} Comparison between experimental \cite{ENSDF} and shell-model calculated low-lying states in $^{96}$Zr and $^{96}$Mo.}
	\end{figure}

	Figure \ref{96Zr_spectra} shows a comparison between the shell-model predicted and the experimental low-lying energy states in $^{96}$Zr and $^{96}$Mo. The $0^+_{\rm g.s.}$ in $^{96}$Zr arises primarily due to the pure $\pi(g_{9/2}^2)\otimes \nu(d_{5/2}^6)$ configuration, whereas the $0^+_2$ and $2^+_1$ states stem from the proton-aligned configuration $\pi(g_{9/2}^2)$ and the collective contribution $\nu(d_{5/2}^4g_{7/2}^2)$ of neutron orbitals, although the wave functions of both states are quite fragmented. Experimentally, the nucleus $^{96}$Zr exhibits a small $B(E2;2^+_1\rightarrow 0^+_1)$ transition strength, verifying the similar value obtained in our calculation. Structure change between the involved initial and final states may cause this small $B(E2)$ value. Similarly, we obtain different structures in the largest configuration of the $8^+_1$ and $6^+_4$ states, namely $[\pi(g_{9/2}^2)\otimes\nu(g_{7/2}^2d_{5/2}^4)]$ and $[\pi(p_{1/2}^2)\otimes\nu(g_{7/2}^1d_{5/2}^5)]$, due to which we record a small $B(E2;8^+_1\rightarrow 6^+_4)$ value, similar to the experimental one (see Table \ref{BE2}). In the case of $^{96}$Mo, the g.s. sequence $0^+_1-8^+_1$ shows similar structure with the dominant configuration $\pi(g_{9/2}^2)\otimes \nu(d_{5/2}^4)$ predicted by the shell model. Thus, as reported in Table \ref{BE2}, the obtained $B(E2)$ values are not weak for the $2^+_1\rightarrow 0^+_1$, $4^+_1\rightarrow 2^+_1$, and $6^+_1\rightarrow 4^+_1$ transitions but still they are not even half of the experimental values. The choice of effective charges might be a reason for relatively smaller $B(E2)$ values, but this is outside the scope of the present study. However, overall, the computed $B(E2)$ values of Table \ref{BE2} are in a reasonable agreement with the experimental data.
 
    After analyzing these results for energy spectra and transition probabilities, we are confident that the present shell-model Hamiltonian can be safely employed in calculating the NMEs ($M_{2\nu}$) for our $2\nu\beta\beta$-decay study.

        \begin{figure}
		\includegraphics[width=87mm]{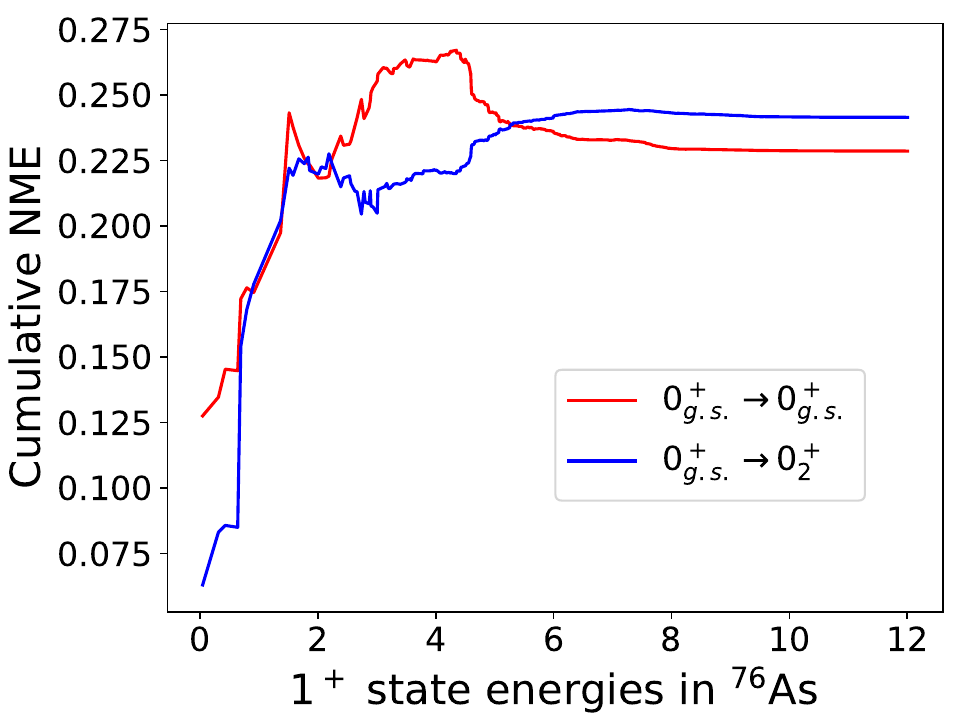}
		\includegraphics[width=87mm]{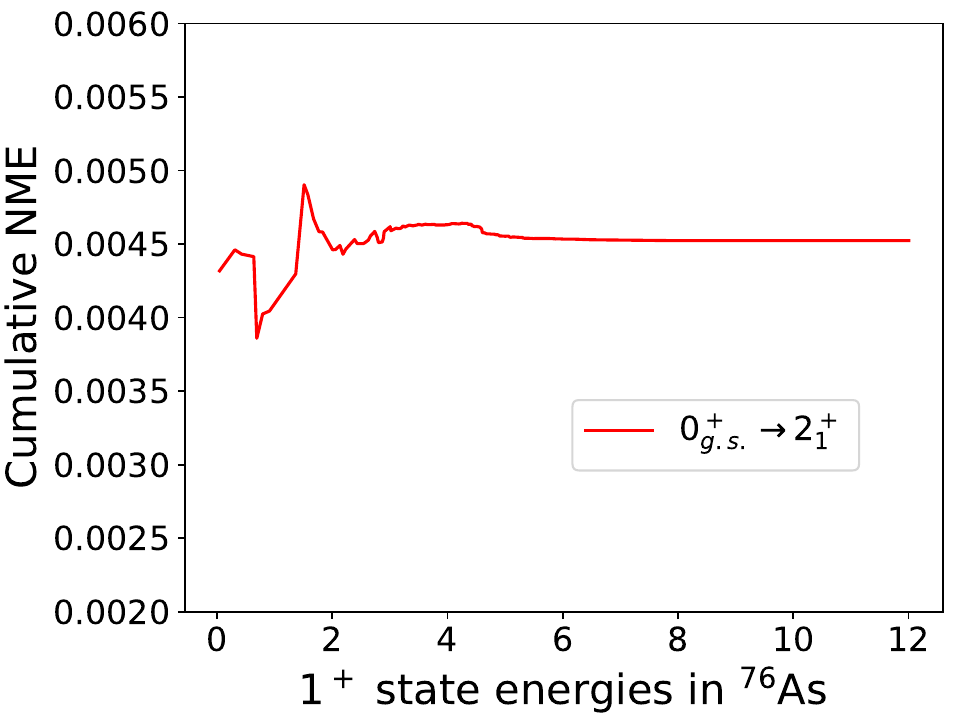}
		
		\caption{\label{matrix1} Cumulative $2\nu\beta\beta$ NMEs ($M_{2\nu}$) for $^{76}$Ge as functions of excitation energy of the intermediate $1^+$ state in $^{76}$As.}
	\end{figure}

	\begin{table*}
		\centering
		\caption{Shell-model calculated $2\nu\beta\beta$ NMEs, extracted effective $g_A$ values, and the experimental half-lives. The phase-space factors $(G^{2\nu})$ are taken from Ref. \cite{Neacsu}.}
		\begin{ruledtabular}
			\begin{tabular}{ccccccc}
				Transition	& $|M_{2\nu}|$ & $G^{2\nu}$ (yr$^{-1}$)  & $g_{\rm A}^{\rm eff}$ &   \makecell{Experimental/Recommended \\ value of $T_{1/2}^{2\nu}$ (yr)}  \\
				\hline
				&  &   &  &  \\
				
				$^{76}\text{Ge}(0^+_{\rm g.s.})\rightarrow$$^{76}\text{Se}(0^+_{\rm g.s.})$	& 0.2285  & 4.51$\times 10^{-20}$ & $0.677\pm 0.004$  & $(2.022\pm{0.018}_{\text{stat}}\pm{0.038}_{\text{syst}})\times10^{21}$ \cite{Agostini}  \\
				
				$^{76}\text{Ge}(0^+_{\rm g.s.})\rightarrow$$^{76}\text{Se}(0^+_2)$	& 0.2414  & 6.4$\times 10^{-23}$ &   & $>6.2\times 10^{21}$ \cite{Klimenko}  \\
				
				$^{76}\text{Ge}(0^+_{\rm g.s.})\rightarrow$$^{76}\text{Se}(2^+_1)$	& 0.0045  & 4.0$\times 10^{-22}$ &  & $>1.1\times 10^{21}$ \cite{Barabash1}  \\
				
			    \hline

                    &  &   & &  \\

                    $^{96}\text{Zr}(0^+_{\rm g.s.})\rightarrow$$^{96}\text{Mo}(0^+_{\rm g.s.})$	& 0.0599  & 642.0$\times 10^{-20}$ & $1.17^{+0.02}_{-0.03}$  & $(2.35\pm 0.14(\text{stat})\pm 0.16(\text{syst}))\times 10^{19}$ \cite{Argyriades} \\

                    $^{96}\text{Zr}(0^+_{\rm g.s.})\rightarrow$$^{96}\text{Mo}(0^+_{2})$	& 0.0159  & 1633.8$\times 10^{-22}$ &   & $>3.1\times 10^{20}$ \cite{Finch} \\

                    $^{96}\text{Zr}(0^+_{\rm g.s.})\rightarrow$$^{96}\text{Mo}(2^+_{1})$	& 7.4$\times 10^{-4}$  & 730.8$\times 10^{-21}$ &  & $>4.1\times 10^{19}$ \cite{Arpesella}  \\

			\end{tabular}
		\end{ruledtabular}
		\label{half-life}
	\end{table*}

	\subsection{Cumulative nuclear matrix elements and the extracted half-lives}
	
\begin{figure}
		\includegraphics[width=87mm]{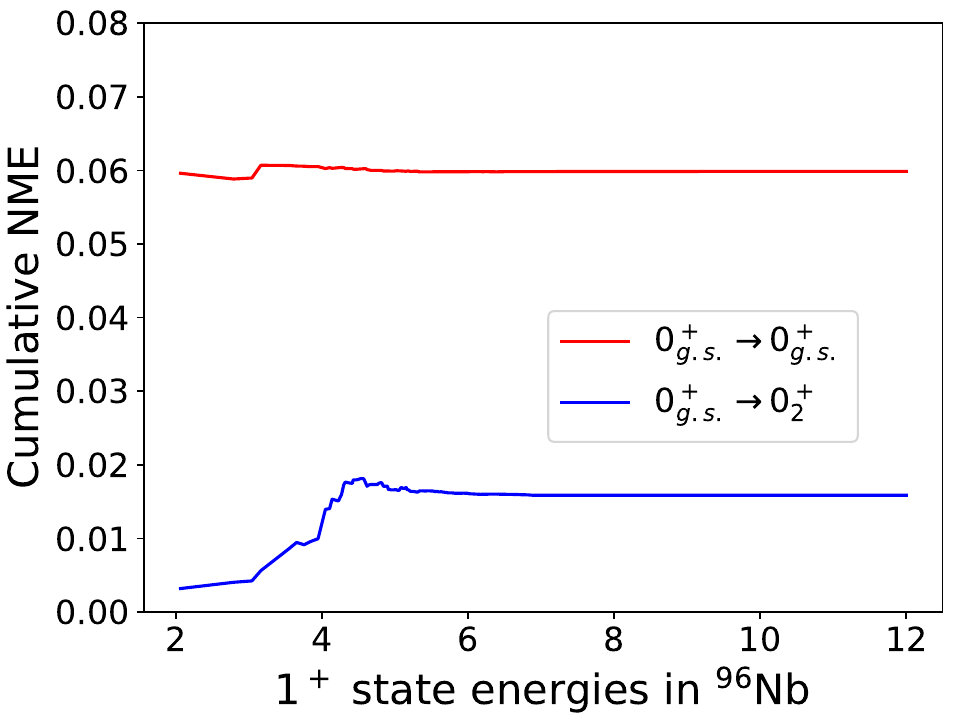}
            \includegraphics[width=87mm]{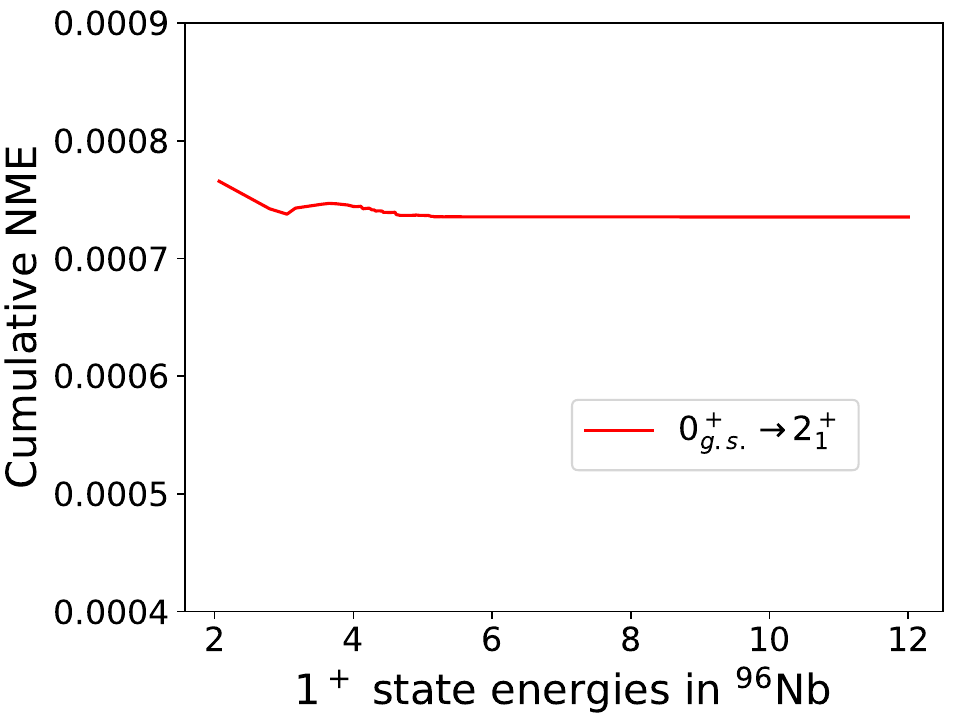}
		
		\caption{\label{96Zr} Cumulative $2\nu\beta\beta$ NMEs ($M_{2\nu}$) for $^{96}$Zr as functions of excitation energy of the intermediate $1^+$ state in $^{96}$Nb.}
	\end{figure}

	Here, we discuss the shell-model calculated results for NMEs and half-lives for the $2\nu\beta\beta$ decays of $^{76}$Ge and $^{96}$Zr. The calculated NMEs and half-lives are reported in Table \ref{half-life}. We have taken the $Q_{\beta\beta}$ and $Q_{\beta}$ values from Ref. \cite{NNDC_qcal}. In the case of $0^+_{\rm g.s.}\rightarrow 0^+_2$ and $0^+_{\rm g.s.}\rightarrow 2^+_1$ transitions, the $Q_{\beta\beta}$ values are taken as the difference of the $Q_{\beta\beta}$ value and the experimental energies of $0^+_2$ and $2^+_1$ states, respectively. The variation of cumulative NMEs as a function of $1^+$ state energies of the intermediate nucleus is depicted in Figs. \ref{matrix1} and \ref{96Zr}.

\begin{table}
		\centering
		\caption{Contribution in the final NME from the first $1^+$ state of the intermediate nucleus.}
		\begin{ruledtabular}
			\begin{tabular}{ccc}
				Transitions	& Value & Percentage (\%)  \\
				\hline
				&  &   \\
				
				$^{76}\text{Ge}(0^+_{\rm g.s.})\rightarrow$$^{76}\text{Se}(0^+_{\rm g.s.})$	& 0.1275  & 55.8   \\

                    $^{76}\text{Ge}(0^+_{\rm g.s.})\rightarrow$$^{76}\text{Se}(0^+_{2})$	& 0.0630 & 26.1   \\

                    $^{76}\text{Ge}(0^+_{\rm g.s.})\rightarrow$$^{76}\text{Se}(2^+_{1})$	& 0.0043 & 95.6   \\

                    \hline
				&  &   \\

                    $^{96}\text{Zr}(0^+_{\rm g.s.})\rightarrow$$^{96}\text{Mo}(0^+_{\rm g.s.})$	& 0.0596  & 99.5 \\

                    $^{96}\text{Zr}(0^+_{\rm g.s.})\rightarrow$$^{96}\text{Mo}(0^+_{2})$	& 0.0032  & 20.1 \\


			\end{tabular}
		\end{ruledtabular}
		\label{first_contribution}
	\end{table}

	{\bf $^{76}$Ge:} A careful and precise calculation of NMEs ($M_{2\nu}$) holds significant importance in the theoretical analysis of the $2\nu\beta\beta$ decay of a nucleus. In the case of $^{76}$Ge, we have calculated $M_{2\nu}$ for three transitions: $0^+_{
\rm g.s.}\rightarrow 0^+_{\rm g.s.}$, $0^+_{\rm g.s.}\rightarrow 0^+_2$, and $0^+_{\rm g.s.}\rightarrow 2^+_1$. In the present calculation, the excitation energies of $1^+$ states in $^{76}$As were shifted such that the lowest-lying $1^+$ state is at the experimental energy of 0.044 MeV. 
As reported in Table \ref{first_contribution}, the first $1^+$ state of the intermediate nucleus contributes 55.8\% of the final NME (0.2285), yielding a value of 0.1275 for the
$0^+_{\rm g.s.}\rightarrow 0^+_{\rm g.s.}$ transition. In this case, we can see from Fig. \ref{matrix1} that the cumulative NME peaks around 4.35 MeV, with a value of $0.2671$. As mentioned earlier, the $pn$QRPA calculations realistically incorporate the contribution of the GTGR region to the NMEs \cite{Suhonen1, Terasaki}. But, to the best of our knowledge, previous shell-model calculations were able to describe this region partly in the case of only $^{48}$Ca \cite{Kostensalo1, Horoi}. In our case for the $0^+_{\rm g.s.}\rightarrow 0^+_{
\rm g.s.}$ transition, the peak is found at the $85$th intermediate $1^+$ state; however, this energy (4.35 MeV) is notably below the energy of the GT resonance ($\approx 11$ MeV) \cite{Thies}. Comparison of our results with those of the large-scale shell-model calculations of Kostensalo \textit{et al.} \cite{Kostensalo} shows that the calculation of \cite{Kostensalo} does not show a prominent peak at this energy, and only a small bump at around 5 MeV of excitation can be discerned. Although, still a large single-particle space is needed in order to reach the collectivity of the GTGR. Near around 12 MeV, the cumulative NMEs for the $0^+_{\rm g.s.}\rightarrow 0^+_{\rm g.s.}$ transition start to saturate and show almost constant behavior with a final value of 0.2285. Similarly, for the $0^+_{\rm g.s.}\rightarrow 0^+_2$ and $0^+_{\rm g.s.}\rightarrow2^+_1$ transitions, the cumulative NMEs nearly saturate after around 11.69 and 7.55 MeV with the final values 0.2414 and 0.0045, respectively. The effective value of the axial-vector coupling strength ($g_{\rm A}^{\rm eff}$) is used to account for the quenching of GT transition strengths. We have extracted the $g_{\rm A}^{\rm eff}$ value for $0^+_{\rm g.s.}\rightarrow 0^+_{\rm g.s.}$ transition using the $G^{2\nu}$, shell-model predicted $M_{2\nu}$, and the experimental half-life \cite{Agostini} similar to the previous study \cite{Kostensalo1}. The total `$\pm$' error ($\sigma$) in the experimental half-life is considered as $\sigma=\sqrt{\sigma_1^2+\sigma_2^2}$, where $\sigma_1$ and $\sigma_2$ are the statistical and systematic errors, respectively. We found the extracted $g_{\rm A}^{\rm eff}$ value as $0.677\pm 0.004$, which is consistent with the previously extracted values reported in \cite{Suhonen}. In the case of $0^+_{\rm g.s.}\rightarrow0^+_2$, and $0^+_{\rm g.s.}\rightarrow2^+_1$ transitions, the cumulative NMEs show a very similar behavior as recorded in the shell-model calculation of \cite{Kostensalo}. The corresponding calculated half-lives using $g_{\rm A}^{\rm eff}$ are $1.276\times 10^{24}$ and $5.877\times 10^{26}$ yr, respectively. Both estimates are notably higher than the experimental lower limits $6.2\times 10^{21}$ \cite{Klimenko} and $1.1\times 10^{21}$ yr \cite{Barabash1}. These results may be useful for experimentalists in estimating the sensitivities of their experimental set-ups to these $0^+_{\rm g.s.}\rightarrow 0^+_2$, and $0^+_{\rm g.s.}\rightarrow 2^+_1$ transitions in the future.

	{\bf $^{96}$Zr:} Because of the large $Q$ value (3356.03 keV), $^{96}$Mo has several excited states through which $^{96}$Zr could undergo $2\nu\beta\beta$ decay. Here, we have taken three transitions $0^+_{\rm g.s.}\rightarrow 0^+_{\rm g.s.}$, $0^+_{\rm g.s.}\rightarrow 0^+_2$, and $0^+_{\rm g.s.}\rightarrow 2^+_1$ from the parent to grand-daughter nucleus for calculating the $M_{2\nu}$ and extracting the $2\nu\beta\beta$-decay half-lives, like in the case of $^{76}$Ge. There is no $1^+$ state having been experimentally confirmed for $^{96}$Nb. Thus, shell-model computed excitation energies are used to calculate the NMEs. In our calculation, we found that the contribution in the cumulative NME from the first $1^+$ state of the intermediate nucleus for the $0^+_{\rm g.s.}\rightarrow 0^+_{\rm g.s.}$ transition is 0.0596, accounting for 99.5\% of the final value (0.0599). Thus, it demonstrates a clear single-state dominance (SSD), consistent with the previous study \cite{Kostensalo1}. On the contrary, the $1^+_1$ state of $^{96}$Nb contributes 20.1\% to the final NME (0.0159) for the $0^+_{\rm g.s.}\rightarrow 0^+_2$ transition. Interestingly, the $0^+_{\rm g.s.}\rightarrow 2^+_1$ transition exhibits SSD with a value 0.00077 of the cumulative NME, which is slightly higher than the final NME (0.00074). Therefore, we have not reported its contributive percentage in Table \ref{first_contribution}. Similar to the $^{76}$Ge, we extracted the $g_{\rm A}^{\rm eff}=1.17^{+0.02}_{-0.03}$ for the $^{96}$Zr$(0^+_{\rm g.s.})\rightarrow^{96}$Mo$(0^+_{\rm g.s.})$ transition using the experimental half-life \cite{Argyriades}. This $g_{\rm A}^{\rm eff}$ is close to the value proposed in Ref. \cite{Kostensalo1}. The extracted half-lives for the $0^+_{\rm g.s.}\rightarrow 0^+_2$, and $0^+_{\rm g.s.}\rightarrow 2^+_1$ transitions corresponding to $g_{\rm A}^{\rm eff}=1.17$ are $1.292 \times 10^{22}$, and $1.334 \times 10^{24}$ yr, respectively. The computed half-lives for the $0^+_{\rm g.s.}\rightarrow 0^+_2$ and $0^+_{\rm g.s.}\rightarrow 2^+_1$ transitions are notably higher than the lower limit of experimental values, which could be useful in estimating experimental sensitivities to these transitions in the future.

 \begin{table*}
		\centering
		\caption{Comparison between our calculated and previously obtained absolute values of NMEs. Here, SM denotes the nuclear shell model, IBM the interacting boson model, HFB the Hartree-Fock-Bogoliubov theory, and (R)QRPA the (renormalized) quasiparticle random-phase approximation. For more values of NMEs of older calculations, see \cite{Suh1998}.}
		\begin{ruledtabular}
			\begin{tabular}{ccccccc}
				
				Transitions	& \multicolumn{6}{c}{ $\vert$ NME $\vert$}  \\

                    \hline
                    
				& Present work & SM \cite{Kostensalo} & IBM \cite{Nomura} & HFB \cite{Dhiman} & RQRPA \cite{Toi1997} & QRPA \cite{Stoica1}  \\

                    \cline{2-7}
				
				$^{76}\text{Ge}(0^+_{\rm g.s.})\rightarrow$$^{76}\text{Se}(0^+_{\rm g.s.})$	& 0.2285  & 0.168  & 0.034 & 0.879 & 0.074 & - \\

                    $^{76}\text{Ge}(0^+_{\rm g.s.})\rightarrow$$^{76}\text{Se}(0^+_{2})$	& 0.2414  & 0.121  & 0.078 & - & 0.130 & - \\

                    $^{76}\text{Ge}(0^+_{\rm g.s.})\rightarrow$$^{76}\text{Se}(2^+_{1})$	& 0.0045  & 0.0012  & - & 1.04 & 0.003 & - \\

                    \hline

                    &  &  &  &  &  \\

                    $^{96}\text{Zr}(0^+_{\rm g.s.})\rightarrow$$^{96}\text{Mo}(0^+_{\rm g.s.})$	& 0.0599  & - & 0.154 & - & 0.036 & 0.022 \\

                    $^{96}\text{Zr}(0^+_{\rm g.s.})\rightarrow$$^{96}\text{Mo}(0^+_{2})$	& 0.0159  & - & 0.063 & - & 0.028 & 0.012 \\

                    $^{96}\text{Zr}(0^+_{\rm g.s.})\rightarrow$$^{96}\text{Mo}(2^+_{1})$	& 7.4$\times 10^{-4}$  & - & - & - & 0.010 & 8.1$\times 10^{-5}$ \\

			\end{tabular}
		\end{ruledtabular}
		\label{NME_comparison}
	\end{table*}

	\subsection{Comparison of NMEs} \label{Comparison_of_NME}

        Here, we discuss the comparison between our shell-model predicted and earlier suggested NMEs using different nuclear models, as reported in Table \ref{NME_comparison}. The NMEs for $2\nu\beta\beta$ decay of $^{76}$Ge have been extracted in different studies \cite{Kostensalo, Nomura, Dhiman, Toi1997, Caurier, Brown, Agostini1, Coraggio, Coraggio1}. We notice that the previously determined NMEs for $^{76}$Ge exhibit either a notably small value (0.034) \cite{Nomura} or significantly large value (0.879) \cite{Dhiman} with respect to correctly reproducing the experimental half-life for the $0^+_{\rm g.s.}\rightarrow 0^+_{\rm g.s}$ transition. The NME (0.168) obtained in Ref. \cite{Kostensalo} is approximately 1.36 times smaller than our calculated NME, and the one of \cite{Toi1997} (0.074) is about half of the one of \cite{Kostensalo}. We have allowed $1p-1h$ excitations across the neutron-shell closure $N=50$, which is not considered in the previous study \cite{Kostensalo}. It is a possible cause for the relatively larger value of the NME in our case. We have accounted for the contribution of a huge number of excited $1^+$ states in the intermediate nucleus $^{76}$As, hence, it is reasonable to involve also the contribution of those $1^+$ states which are arising due to the $1p-1h$ excitations across the $N=50$ shell gap. We can see that some spin-orbit partner orbitals are absent in our model space, particularly neutron orbital $\nu(1p_{3/2})$. As a consequence, no contribution of GT transition is possible via $\nu(p_{3/2})\rightarrow \pi(p_{1/2})$ and $\nu(p_{3/2})\rightarrow \pi(p_{3/2})$ paths. However, we have considered the primary neutron-proton orbitals that contribute to GT transitions, e.g., the most significant GT transitions in the $2\nu\beta\beta$ decay of $^{76}$Ge can occur via $p_{1/2}\rightarrow p_{1/2}$ and $g_{9/2}\rightarrow g_{9/2}$ paths. For those $1^+$ states of the intermediate nucleus that lie at higher energies, the GT transitions will be driven by $\nu(g_{7/2})\rightarrow \pi(g_{9/2})$ path. We tested the Ikeda sum rule for $^{76}$Ge and $^{76}$Se to assess the efficiency of the used model space, which is not discussed in previous studies \cite{Kostensalo, Dhiman, Caurier}. We found that the difference of total GT strengths $\sum B(GT^-)$ and $\sum B(GT^+)$ considering 5000 $1^+$ states in $^{76}$As and $^{76}$Ga for $^{76}$Ge$(0^+_{\rm g.s.})\rightarrow^{76}$As$(1^+)$ and $^{76}$Ge$(0^+_{\rm g.s.})\rightarrow^{76}$Ga$(1^+)$ transitions is 6.047. Thus, we can obtain the effective Ikeda sum rule (EISR) $3(N_{\rm core}-Z_{\rm core})+\sum B(GT^-)-\sum B(GT^+)$ value as $(3\times10)+6.047=36.047$. Here, $N_{\rm core}$ and $Z_{\rm core}$ are the number of neutrons and protons in the core nucleus $^{66}_{28}$Ni$_{38}$. The above EISR value stems from the relation $3(N_{\rm act}-Z_{\rm act})=\sum B(GT^-)-\sum B(GT^+)$ based on the discussion in section (17.4.3), on page 545 of Ref. \cite{Suh2007}. Here, $N_{\rm act}$ and $Z_{\rm act}$ are the active number of neutrons and protons outside the core. In the case of $^{76}$Se, the calculated difference of $\sum B(GT^-)$ and $\sum B(GT^+)$ considering 5000 $1^+$ states in $^{76}$Br and $^{76}$As for $^{76}$Se$(0^+_{\rm g.s.})\rightarrow^{76}$Br$(1^+)$ and $^{76}$Se$(0^+_{\rm g.s.})\rightarrow^{76}$As$(1^+)$ transitions is $-1.875$, and the computed EISR value is $(3\times10)-1.875=28.125$. Experimentally, we expect the EISR value $3(N-Z)$ for $_{32}^{76}$Ge$_{44}$ and $_{34}^{76}$Se$_{42}$ as 36 and 24, respectively. Therefore, one can say that the EISR is practically exact in the case of $^{76}$Ge but shows some deviation for $^{76}$Se. This deviation occurs significantly due to the absence of $\nu(1p_{3/2})$ and $\nu(0f_{5/2})$ orbitals in our model space. For both the proton and neutron model spaces, the absence of spin-orbit partner $0f_{7/2}$ of $0f_{5/2}$ orbital can also make a small effect on the calculated $2\nu\beta\beta$ NMEs.

        Notably, the final NME for the $0^+_{\rm g.s.}\rightarrow 0^+_2$ transition is larger than for the ground-state transition. Similar results concerning the calculated NMEs for the $0^+_{\rm g.s.}\rightarrow 0^+_{\rm g.s.}$ and $0^+_{\rm g.s.}\rightarrow 0^+_{2}$ transitions were also found in previous studies, utilizing the interacting boson model (IBM) \cite{Nomura} and the renormalized quasiparticle random-phase approximation (RQRPA) \cite{Toi1997} for $^{76}$Ge, as also the quasiparticle random-phase approximation (QRPA) \cite{Jokiniemi} for $^{136}$Xe.

 Previously, \v{S}imkovic \textit{et al.} \cite{Simkovic} proposed that the NME for the $0^+_{\rm g.s.}\rightarrow 0^+_{\rm g.s.}$ transition in $2\nu\beta\beta$ decay of $^{76}$Ge is suppressed because of the different deformations of the parent and granddaughter nucleus. We can deduce the deformation parameter $(\beta_2)$ from the intrinsic quadrupole moment ($Q_0$). The $\beta_2$ value is related to the $Q_0$ moment by the following expression \cite{Kilgallon}

        \begin{equation} \label{eq4}
            \beta_2=\frac{\sqrt{5 \pi}}{3ZR^2}Q_0.
        \end{equation}
Here, for simplicity, we have taken $R$ as $1.2\times A^{1/3}$ fm. Initially, we determined the $Q_0$ moment using the spectroscopic quadrupole moment, $Q_s(2^+_1)=-28.714$ $e$fm$^2$ in $^{76}$Ge obtained from our shell-model calculation, employing the formula $Q_0=-(7/2)Q_s$ \cite{Simkovic} which arises from the expression \cite{Kilgallon}

\begin{equation}
	Q_s=\frac{3K^2-I(I+1)}{(I+1)(2I+3)}Q_0,
\end{equation}

 by taking $K=0$ for $I^{\pi}=2^+$. Substituting the $Q_0$ value into Eq. (\ref{eq4}), we find the $\beta_2$ value to be approximately 0.161. Similarly, utilizing the calculated $Q_s(2^+_1)$ value ($37.142$ $e$fm$^2$) for $^{76}$Se from our shell-model calculation, we obtain the value $\beta_2=-0.195$. Previous $\beta_2$ values computed for $^{76}$Ge (0.157) and $^{76}$Se ($-0.244$) within relativistic mean field (RMF) theory \cite{Lalazissis} also support our findings. As suggested in Ref. \cite{Simkovic}, the behavior of final NME can be predicted by the value $\Delta \beta_2=|\beta_2(^{76}\text{Ge})-\beta_2(^{76}\text{Se})|$ and the suppression of the $2\nu\beta\beta$-decay NME becomes stronger with an increase of $\Delta \beta_2$. Different deformation (prolate for $^{76}$Ge and oblate for $^{76}$Se) yields a larger $\Delta \beta_2$, which may be a possible factor for the suppression of the NME for the $0^+_{\rm g.s.}\rightarrow 0^+_{\rm g.s.}$ transition in our case (see Fig. \ref{matrix1}).


        In the case of $^{96}$Zr, our computed value of the NME, 0.0599, for the $0^+_{\rm g.s.}\rightarrow 0^+_{\rm g.s.}$ transition, is slightly smaller than the 0.0747 one of \cite{Kostensalo1}. Looking at Table \ref{NME_comparison}, the previously determined NME using the IBM is larger than those obtained in the other theory frameworks. In fact, the present shell-model results for this transition are in good agreement with the  RQRPA results of Toivanen \textit{et al.} \cite{Toi1997} and QRPA results of Stoica \cite{Stoica1}. For the $0^+_{\rm g.s.}\rightarrow 0^+_2$ transition, a similar situation prevails, the IBM result deviating notably from the rest. Concerning the $0^+_{\rm g.s.}\rightarrow 2^+_{1}$ transition, the RQRPA \cite{Toi1997} gives a much larger NME than the present calculation and the QRPA model of Stoica \cite{Stoica1}. Our NME for the $0^+_{\rm g.s.}\rightarrow 2^+_{1}$ transition agrees qualitatively with the one of \cite{Stoica1} although is one order of magnitude larger than it. Here, it has to be pointed out that we have completely blocked neutron excitations across the $N=50$ shell closure in order to make the calculation of the NMEs for $^{96}$Zr feasible due to which there is only $\nu(g_{7/2})\rightarrow \pi(g_{9/2})$ decay path possible for GT transitions. Further, restricting the occupancy of the $\pi(0g_{9/2})$ orbital to a maximum of four protons may cause inaccuracies in predicting the wave functions of the states of interest. It also reflects in the calculated difference of $\sum B(GT^-)$ and $\sum B(GT^+)$ as $-0.3024$, for the transitions $^{96}$Zr$(0^+_{\rm g.s.}) \rightarrow ^{96}$Nb$(1^+)$ and $^{96}$Zr$(0^+_{\rm g.s.}) \rightarrow ^{96}$Y$(1^+)$, and the EISR shows large deviation from our calculated result. Here, the absence of neutron orbitals $\nu(0f_{5/2}1p0g_{9/2})$ is significantly responsible for this discrepancy. However, it is not possible to test the Ikeda sum rule for $^{96}$Mo$(0^+_{\rm g.s.})\rightarrow ^{96}$Tc$(1^+)$ and $^{96}$Mo$(0^+_{\rm g.s.})\rightarrow ^{96}$Nb$(1^+)$ transitions due to the model space truncation. In the future, one could achieve more precise NMEs if it was be possible to include additional proton excitations in $\pi(0g_{9/2})$ orbital and the contribution of the $\nu(1p_{1/2})$ and $\nu(0g_{9/2})$ orbitals.

 \begin{table*}
	\centering
	\caption{Comparison of shell-model calculated level density (number of states in the per MeV interval) of $1^+$ states in $^{76}$As and $^{96}$Nb with the back-shifted Fermi gas model. The level density using the formula from back-shifted Fermi gas model is calculated utilizing the endpoint energies of the interval [e.g., for the interval ($0-1$), we have taken $E_x=1$ MeV].}
	\begin{ruledtabular}
		\begin{tabular}{ccccccc}
			
			\makecell{Shell-model energy \\ interval (in MeV)}	& \multicolumn{2}{c}{ \makecell{Level density in $^{76}$As \\ (number of states/MeV)}} & \multicolumn{2}{c}{ \makecell{Level density in $^{96}$Nb \\ (number of states/MeV)}} \\
			
			\hline
			
			& Shell-model & \makecell{Back-shifted \\ Fermi gas model} & Shell-model & \makecell{Back-shifted \\ Fermi gas model}  \\
			
			\cline{2-3}
                \cline{4-5}
			
			$0-1$	& 7   & 2.859 & 0 & 4.889 \\
			
			$1-2$	& 7   & 6.379 & 0 & 12.339 \\
			
			$2-3$	& 18  & 13.192 & 2 & 28.641 \\
			
			$3-4$	& 34  & 25.853 & 8 & 62.504 \\
			
			$4-5$	& 64  & 48.603 & 32 & 129.952 \\
			
			$5-6$	& 113 & 88.342 & 100 & 259.694 \\
			
			$6-7$	& 177 & 156.111 & 255 & 501.995 \\
			
			$7-8$	& 282 & 269.292 & 549 & 943.104 \\
			
			$8-9$	& 436 & 454.880 & 864 & 1728.389 \\
			
		   $9-10$ & 485 & 754.275 & 568 & 3098.968  \\
		   
		  $10-11$ & 334 & 1230.250 & 460 & 5449.031  \\
		  
		  $11-12$	& 281 & 1976.991 & 408 & 9414.593  \\
		  
			
		\end{tabular}
	\end{ruledtabular}
	\label{level_table}
\end{table*}

\subsection{Level density of $1^+$ states}

         The Bethe formula of the level density \cite{Bethe} from the BFM considering the spin dependence can be written as \cite{Gilbert}

          \begin{equation}
	      \rho(U,J)=\rho(U).f(J).
          \end{equation}

       Here,

           \begin{equation}
	       \rho(U)=\frac{1}{12\sqrt{2}\sigma}\frac{{\rm exp}(2\sqrt{aU})}{a^{1/4}U^{5/4}}
           \end{equation}

       and

           \begin{equation}
              f(J)=\frac{(2J+1){\rm exp}[-(J+\frac{1}{2})^2/2\sigma^2]}{2\sigma^2},
           \end{equation}

       where the spin cutoff parameter \cite{Gilbert}, $\sigma^2=0.0888 A^{2/3} \sqrt{a(E_x-\Delta)}$. The level density parameter `$a$' is given by $a=(A/16)$ MeV$^{-1}$ \cite{Gross}; $U$ is defined as $U=(E_x-\Delta$) \cite{Koning}; $E_x$ is the excitation energy of the spin-state $J$; $A$ is the atomic mass. The back-shifting parameter $\Delta$ (in MeV) can be defined as \cite{Rauscher}

	\begin{equation}
		\Delta=
		\left\{
		\begin{aligned}
			& 12/\sqrt{A} ~~~~~~~~ \text{for even-even nuclei} \\
			& 0 ~~~~~~~~~~~~~~~~ \text{for odd-}A ~ \text{nuclei} \\
			& -12/\sqrt{A} ~~~~ \text{for odd-odd nuclei}.
		\end{aligned}
		\right.
	\end{equation}
	
         Considering the equiparity distribution, the level density can be written as \cite{Koning}

        \begin{equation} \label{parity}
	    \rho(U,J,\pi)=\frac{1}{2}\rho(U).f(J).
        \end{equation}

 We have reported the level density of $1^+$ states in $^{76}$As and $^{96}$Nb from the shell-model calculations and the level density of $1^+$ states for odd-odd nuclei with $A=76$ and $A=96$ from the BFM [using Eq. (\ref{parity})] in Table \ref{level_table}.
 
 It is clear from Table \ref{level_table} that the shell-model calculated level density of $1^+$ states in $^{76}$As is consistent with the level density calculated with back-shifted Fermi gas model up to $9-10$ MeV but beyond this energy shows deviation from the estimated level density with the BFM. Thus, we can say that shell-model calculations consider the behavior of level density more accurately, up to $9-10$ MeV, and after that, deviate due to the model space truncation. In conclusion, concerning the computed NMEs of the various discussed $2\nu\beta\beta$ transitions, it can be stated that for the $^{76}$Ge decay, the cumulative NMEs of Fig. \ref{matrix1} might be poorly approximated by the shell-model after some 8 MeV of excitation energy in $^{76}$As owing to the insufficient number of $1^+$ states above this energy. The same can be said about the NME of $^{96}$Zr, where the number of shell-model computed $1^+$ states is reasonable only up to $7-8$ MeV (see Table \ref{level_table}). Although the shell-model calculated level densities of $1^+$ states in $^{96}$Nb are slightly lower from the starting than that predicted by the BFM, this may be due to the constraints of our model space for the $2\nu\beta\beta$ decay study of $^{96}$Zr (as discussed in Sec. \ref{Comparison_of_NME}) and points to areas for future improvement.


        \subsection{Branching ratios}

         The nucleus $^{76}$Ge can decay via $2\nu\beta\beta$ mode into the g.s. and different excited states of $^{76}$Se. But, the nucleus $^{96}$Zr can decay via $2\nu\beta\beta$ mode into the g.s. and different excited states of $^{96}$Mo and also it can decay via single-$\beta$ transitions into the lowest $4^+$, $5^+$, and $6^+$ states of $^{96}$Nb. Here, we discuss the shell-model predicted branching ratios for the $2\nu\beta\beta$ and/or single-$\beta$ transitions for the $^{76}$Ge and $^{96}$Zr decays corresponding to our estimated $g_{\rm A}^{\rm eff}$ values.

         In the case of $^{76}$Ge, we first calculated the total half-life for the $0^+_{\rm g.s.}\rightarrow 0^+_{\rm g.s.}$, $0^+_{\rm g.s.}\rightarrow 0^+_2$, and $0^+_{\rm g.s.}\rightarrow 2^+_1$ transitions of $2\nu\beta\beta$ decay using the formula \cite{Suh2007}

         \begin{equation}
             \frac{1}{T_{1/2}^{\rm total}}=\sum_k \frac{1}{T_{1/2}^k},
         \end{equation}
         
         where $k$ refers to the final state of the decay. Subsequently, we have calculated the branching ratios (B.R.), $({\rm B.R.})^k=T_{1/2}^{\rm total}/T_{1/2}^k$ \cite{Suh2007} for all three transitions. It is clear from the computed B.R. that the $0^+_{\rm g.s.}\rightarrow 0^+_{\rm g.s.}$ transition is dominant having the largest B.R. of 99.84\%. The B.R. is notably smaller for the $0^+_{\rm g.s.}\rightarrow 0^+_2$ transition (15.81$\times 10^{-2}$\%), and negligible for the $0^+_{\rm g.s.}\rightarrow 2^+_1$ transition (3.43$\times 10^{-4}$\%).

         Our estimated half-lives for the fourth-forbidden non unique, fourth-forbidden unique, and sixth-forbidden non unique $\beta$ transitions from the $0^+_{\rm g.s.}$ of $^{96}$Zr to the $4^+_1$, $5^+_1$, and $6^+_1$ states of $^{96}$Nb (corresponding to $g_{\rm A}=1.17;g_{\rm V}=1.00$) are $9.574\times 10^{22}$, $1.034\times 10^{20}$, and $2.247\times 10^{29}$ yr, respectively. These are quite similar to the half-lives $7.5\times 10^{22}$, $1.1\times 10^{20}$, and $1.6\times 10^{29}$ yr obtained in Ref. \cite{Alanssari}. Our results predict that the $2\nu\beta\beta$ decay branching ratio (81.49\%) is larger than the $\beta$ decay branching ratio (18.51\%) in $^{96}$Zr, consistently with the earlier shell-model study \cite{Kostensalo1} where the $2\nu\beta\beta$ branching ratio was 81.6\%. In the QRPA study of \cite{Hei2007} the $\beta$ decay was dominated by the $0^+_{\rm g.s.}\rightarrow 5^+_1$ transition with a half-life of $2.4\times 10^{20}$ yr, in agreement with the present study. In the $2\nu\beta\beta$ branch of $^{96}$Zr, the $0^+_{\rm g.s.}\rightarrow 0^+_{\rm g.s.}$ transition is dominant with a 81.34\% contribution, while the contributions from $0^+_{\rm g.s.}\rightarrow 0^+_{2}$ (1.48$\times 10^{-1}$\%), and $0^+_{\rm g.s.}\rightarrow 2^+_{1}$ (1.43$\times 10^{-3}$\%) transitions are almost negligible. In the case of the single-$\beta$ branch, the $0^+_{\rm g.s.}\rightarrow 5^+_1$ transition dominates with 18.49\% contribution, consistently with the previous  study \cite{Kostensalo1}. We found significantly smaller branching ratios for the $0^+_{\rm g.s.}\rightarrow 4^+_1$ (0.02\%) and $0^+_{\rm g.s.}\rightarrow 6^+_1$ (0.85$\times 10^{-8}$\%) transitions, again consistently with the studies of Kostensalo \textit{et al.} \cite{Kostensalo1} and Heiskanen \textit{et al.} \cite{Hei2007}.

	\section{Summary and Conclusions} \label{section4}

        In the present study, we have investigated the $2\nu\beta\beta$ decay in $^{76}$Ge and $^{96}$Zr isotopes by using the GWBXG shell-model effective Hamiltonian. First, we validated the applicability of this Hamiltonian by computing low-lying energy levels and reduced transition probabilities in both parent and granddaughter nuclei. Our findings indicate that the present shell-model interaction is suitable for calculating NMEs for $2\nu\beta\beta$ decay. To ensure the convergence of cumulative NMEs, we computed 5000 $1^+$ states in $^{76}$As and $^{96}$Nb. The effective $g_A$ values are estimated for the $0^+_{\rm g.s.}\rightarrow 0^+_{\rm g.s.}$ transitions using shell-model predicted NMEs and measured half-lives for the $2\nu\beta\beta$ decay of both nuclei. Utilizing the calculated NMEs and obtained $g_{\rm A}^{\rm eff}$, we extracted the half-lives of $^{76}$Ge and $^{96}$Zr for the $0^+_{\rm g.s.}\rightarrow 0^+_{2}$, and $0^+_{\rm g.s.}\rightarrow 2^+_{1}$ transitions through $2\nu\beta\beta$ decay. We further investigated the variation of cumulative NMEs with respect to the energies of $1^+$ states of the intermediate nucleus for all three transitions. Our analyses reveal that the present shell-model Hamiltonian partially describes the contribution of the GTGR region in the case of $^{76}$Ge for the $0^+_{\rm g.s.}\rightarrow 0^+_{\rm g.s.}$ transition, which was not obtained in the previous shell-model studies. Our calculations show single-state dominance (SSD) for the $0^+_{\rm g.s.}\rightarrow 0^+_{\rm g.s.}$ transition in $^{96}$Zr. The computed half-lives turned out to be in good agreement with the experimental data. Additionally, we compared the predicted NMEs with those obtained from different nuclear models in previous works. We studied the variation of level density in $1^+$ states of $^{76}$As from the shell-model and back-shifted Fermi gas model. The shell-model predicted branching ratios for the $2\nu\beta\beta$ decay of $^{76}$Ge, and both the single-$\beta$ and $2\nu\beta\beta$ decay of $^{96}$Zr are also analyzed.

        We have tried to include the relevant proton and neutron orbitals in our model space as much as possible with our current computational facilities. But, there are still some issues with the highly excited $1^+$ states of the intermediate nuclei at higher energies where their collectivity could be affected due to the lack of some GT connections between different orbitals, especially in the case of $^{96}$Zr. Here, it also cannot be ignored that the enhancement in the model space is still computationally very challenging due to the huge shell-model dimensions. We will perform these calculations with more enlarged model spaces once advanced computational facilities allow us to do so.

	\section*{\uppercase{Acknowledgements}}

	We acknowledge financial support from SERB (India), Grant No. CRG/2022/005167. Additionally, D.P. acknowledges financial support from MHRD, the Government of India. We would like to thank the National Supercomputing Mission (NSM) for providing computing resources of ``PARAM Ganga'' at the IIT Roorkee, implemented by C-DAC and supported by MeitY and DST, Government of India.

\end{document}